\documentclass{emulateapj}
\usepackage{graphics}
\usepackage{amssymb}
\usepackage{amsmath}
\usepackage[dvips]{color}

\newcommand{\A}{$\rm \AA$}

\newcommand{\feii}{Fe\,{\footnotesize II}}

\newcommand{\mgii}{Mg\,{\footnotesize II}}
\newcommand{\heii}{He\,{\footnotesize II}}
\newcommand{\nv}{N\,{\footnotesize V}}

\newcommand{\oiii}{O\,{\footnotesize III}]}

\newcommand{\civ}{C\,{\footnotesize IV}}

\newcommand{\siiv}{Si\,{\footnotesize IV}}

\newcommand{\etal}{et al.}

\begin{document}

\title{Outflow and hot dust emission in broad absorption line quasars}

\author{ Shaohua Zhang\altaffilmark{1}, Huiyuan Wang\altaffilmark{2}, Tinggui Wang\altaffilmark{2},
Feijun Xing\altaffilmark{2}, Kai Zhang\altaffilmark{3}, Hongyan Zhou\altaffilmark{1,2}, Peng Jiang\altaffilmark{2}}
\altaffiltext{1}{Polar Research Institute of China, 451 Jinqiao Road, Shanghai, 200136, China;
zhangshaohua@pric.gov.cn}
\altaffiltext{2}{Key Laboratory for Research in Galaxies and Cosmology, University of Science and Technology of
China, Chinese Academy of Sciences, Hefei, Anhui, 230026, China; whywang@mail.ustc.edu.cn}
\altaffiltext{3}{Key Laboratory for Research in Galaxies and Cosmology, Shanghai Astronomical Observatory, Chinese
Academy of Sciences, 80 Nandan Road, Shanghai 200030, China}

\begin{abstract}
We have investigated a sample of 2099 broad absorption line (BAL) quasars with $z=1.7-2.2$ built from the Sloan Digital Sky Survey Data Release Seven and the Wide-field Infrared Survey. This sample is collected from two BAL quasar samples in the literature, and refined by our new algorithm. Correlations of outflow velocity and strength with hot dust indicator ($\beta_{\rm NIR}$) and other quasar physical parameters, such as Eddington ratio, luminosity and UV continuum slope, are explored in order to figure out which parameters drive outflows. Here $\beta_{\rm NIR}$ is the near-infrared continuum slope, a good indicator of the amount of hot dust emission relative to accretion disk emission. We confirm previous findings that outflow properties moderately or weakly depends on Eddington ratio, UV slope and luminosity. For the first time, we report moderate and significant correlations of outflow strength and velocity with $\beta_{\rm NIR}$ in BAL quasars.
It is consistent with the behavior of blueshifted broad emission lines in non-BAL quasars. The statistical analysis and composite spectra study both reveal that outflow strength and velocity are more strongly correlated with $\beta_{\rm NIR}$ than Eddington ratio, luminosity and UV slope. In particular, the composites show that the entire \ion{C}{4} absorption profile shifts blueward and broadens as $\beta_{\rm NIR}$ increases, while Eddington ratio and UV slope only affect the high and low velocity part of outflows, respectively.
We discuss several potential processes and suggest that dusty outflow scenario, i.e. dust is intrinsic to outflows and may contribute to the outflow acceleration, is most likely.
The BAL quasar catalog is available from the authors upon request.
\end{abstract}


\section{Introduction}

Outflows appear to be a common phenomenon in quasars. Strong outflows may carry away huge amounts of material, energy and angular momentum and are believed to be one of the most important feedback processes connecting their central engines and host galaxies (e.g. Hopkins \& Hernquist 2006; Hopinks \& Elvis 2010). For instance, outflows are thought to be able to regulate the growth of supermassive black holes (SMBH) and star formation in host galaxies (see Antonuccio-Delogu \& Silk 2010 for a recent review), chemically enrich the interstellar medium (ISM) of host galaxies and the surrounding intergalactic medium (IGM) (e.g., Collin \& Zahn 1999; Veilleux et al. 2005). Lots of studies tried to closely tie up outflows with the quasar fundamental parameters, such as Eddington ratio (Boroson 2002; Gangly et al. 2007; Zhang et al. 2010; Wang et al. 2011; Marziani \& Sulentic 2012), the intrinsic spectral energy distribution (SED, e.g. Leighly \& Moore
2004; Fan et al. 2009; Richards et al. 2011; Baskin et al. 2013) and even gas metallicity (e.g. Wang et al. 2012). Therefore outflows are important for our understanding of the evolution of the central SMBHs and the connection with host galaxies.

Quasar outflows manifest themselves in blueshifted absorption and emission, such as broad absorption lines (BALs; Weymann et al. 1991) and blueshifted broad emission lines (BELs, Gaskell 1982). It is well known that BAL quasars in general have redder ultraviolet (UV) continua than non-BAL quasars, and quasars with BALs from low-ionization state species (e.g. \mgii\ and \feii\; LoBALs) are even redder than quasars with absorption in high-ionization species (e.g., \civ\ and \nv; HiBALs) on average (Weymann et al. 1991; Brotherton et al. 2001; Reichard et al. 2003). This phenomenon is readily interpreted as a result of dust extinction (e.g. Sprayberry \& Foltz 1992; Voit et al. 1993; Reichard et al. 2003; Hewett \& Foltz 2003; Dai et al. 2008; Jiang et al. 2013). The BAL outflow launching region is suggested to be co-spatial with or outside of the BEL regions, based on the observational fact that BAL features usually obscure BELs. Dust can survive at the outer boundary of BEL regions according to the reverberation mapping results for local active galactic nuclei (Suganuma et al. 2006). It implies the possibility of dusty outflows. Recently, Grupe et al. (2013) found an interesting anti-correlation between luminosity and UV continuum slope in a variable BAL Seyfert 1, WPVS 007. It also favors a dusty outflow component moving transversely across our line of sight.

The composite spectra studies suggested an SMC-like reddening law for dust associated with quasar outflows (Reichard et al. 2003; Zhang et al. 2010), though it may not always be the case (see the studies of some individual objects, e.g. Hall et al. 2002; Jiang et al. 2013). The average amount of reddening in HiBAL and LoBAL quasars are $E(B-V)\sim$0.023 and $E(B-V)\sim$0.077, respectively (Reichard et al. 2003; Zhang et al. 2010). Such large amount of radiation absorbed/scattered by dust grains must be partly re-radiated in infrared band, yielding a prominent and detectable feature. More recently, Wang et al. (2013, hereafter Paper I) found an unexpected correlation between the amount of hot dust emission relative to accretion disk emission and the blueshift of \civ\ BEL in $z\sim2$ non-BAL quasars. The correlation dramatically strengthens with increasing Eddington ratio, since outflows tend to be dominant in high Eddington ratio quasars. It strongly implies an important role of dust in the outflow physics, such acceleration mechanisms or interaction with nearby medium (Paper I), and even in the galaxy physics and cosmology (see Elvis et al. 2002).

BALs and blueshifted BELs are very likely to be different appearances of the same outflow component viewed from different inclination
angles (see e.g. Wang et al. 2011).  The normal BELs emitted by virialized (or rotational) gases may hamper the proper measurement of outflow properties from BELs. Fortunately this effect is negligible in BAL measurements. It is thus necessary to revisit and extend the relationship between outflow properties and hot dust emission using a BAL quasar sample. In this paper, we construct a large $z\sim2$ BAL quasar sample from the Wide-field Infrared Survey Explorer (WISE; Wright et al. 2010) and the Sloan Digital Sky Survey (SDSS; York et al.2000). Following Paper I, we adopt the rest-frame NIR slope, measured from WISE data, as an indicator of the relative amount of hot dust emission. We use the BAL parameters measured from SDSS spectra, such as velocities and absorption strength, to indicate the outflow strength as commonly adopted in the literature.

This paper is organized as follows. The sample construction and outflow parameter measurement are shown in Section 2.
We analyze the data and present the correlations in Section 3. Finally, we discuss the possible underlying physics in Section 4,
and summarize the results in Section 5. Throughout this paper, we adopt the CDM `concordance' cosmology
with H0 = 70 km s$^{-1}$Mpc$^{-1}$, $\Omega_{\rm m} = 0.3$, and $\Omega_{\Lambda} = 0.7$.

\section{Sample Construction}

\figurenum{1}
\begin{figure*}[tbp]
\epsscale{1.0} \plotone{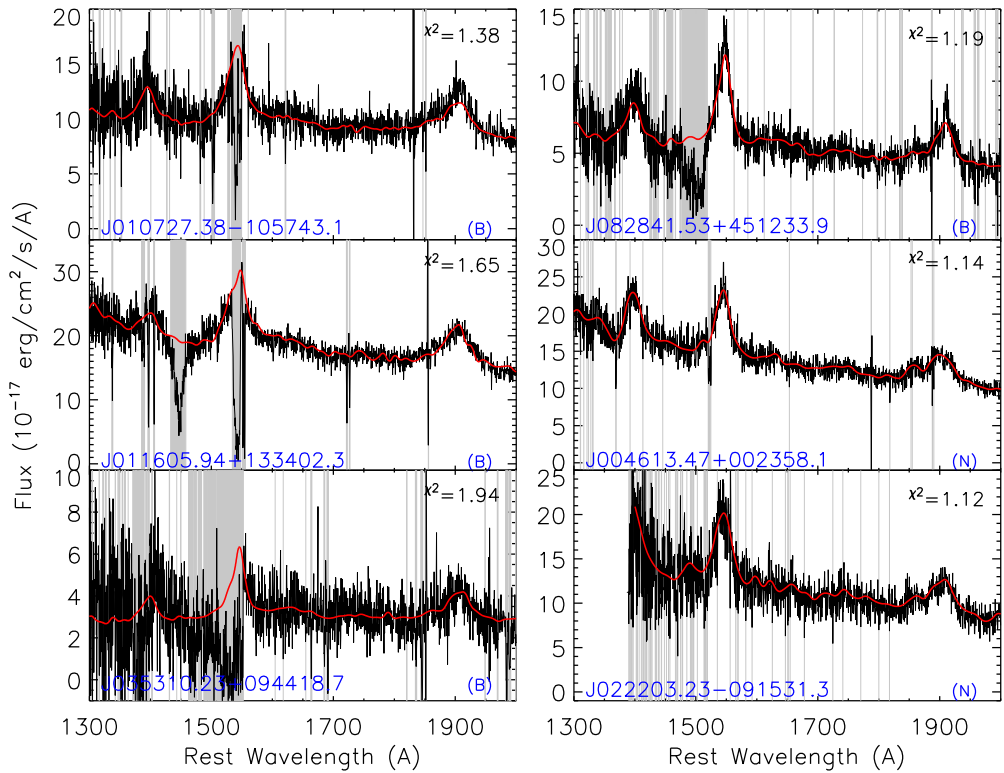}
\caption{Observed spectra (black) overplotted with the corresponding best-fitting model spectra (red) for four BAL quasars (marked with ``B")
and two non-BAL quasars (marked with ``N"). Note that the two non-BAL quasars  had been identified as BAL quasars by other works (see text for details). The gray regions are masked out in the second loop of step 2 and the calculation of $\chi^2$ and $\chi_{\rm EL}^2$. They are
considered to be the potential broad absorption regions or narrow absorption lines (see Appendix A for details). The number in each panel is the reduced $\chi^2$.}\label{fig1}
\end{figure*}

We start from the SDSS DR7 quasar catalog (Schneider et al. 2010), and select 23,059 quasars with redshifts of
$1.7<z<2.2$. The redshift range is chosen to ensure reliable measurements of \ion{C}{4} BALs
and \ion{Mg}{2} BELs, the latter can be used to derive quasars' fundamental parameters.
We coordinate cross-match these quasars with entries from the WISE All Sky Data Release catalog, using a match radius of 3 arcseconds, and get 21,720 objects detected by WISE. WISE surveyed the whole sky at four infrared bands, i.e. W1, W2, W3 and W4. The effective wavelengths of the first three bands are $\lambda_1=3.35$, $\lambda_2=4.60$ and $\lambda_3=11.56$, which covers the rest-frame NIR band (from $1\mu$m to $4\mu$m) for the selected quasars. Since we are interested in the rest-frame NIR emission, we only consider the first three bands in this work.
After discarding the quasars with signal to noise ratio (S/N) less than three in the three WISE bands,
we obtained 18,451 quasars that compose the DR7-WISE quasar catalog.
Then cross-correlating the DR7-WISE quasar catalog with the SDSS DR5
learning vector quantization (LVQ) \ion{C}{4} BAL quasar catalog
(Scaringi et al. 2009, hereafter SC09) and the SDSS DR7 BAL quasar catalog
we compile a BAL quasar candidate catalog, which contains 2964 BAL quasar candidates (these candidates are identified as BAL quasars either by SC09 or SH11).

\subsection{Identification and Measurement of Absorption Troughs}

This candidate catalog is not our final working sample. The BAL quasar candidates are collected from
two different samples in the literature, so the selection criterion in our candidate catalog is not self-consistent. For instance, a considerable amount of candidates are included by one work, but rejected by the other. In addition, only part of candidates have BAL parameter measurements. Therefore it is essential to use the same criterion to refine the sample, and then measure the BAL parameters for the whole sample. We develop a new method to construct our working sample from the candidate catalog.
The details of our algorithm are presented in Appendix A, and here we only briefly describe it.
The basic idea, same as previous works, is to construct an unabsorbed model spectrum (hereafter model spectrum)
which adequately resembles the unabsorbed spectral regions of a given candidate, and then compare it with the given spectrum to identify potential absorption troughs.

For a given BAL quasar candidate, the model spectrum is obtained through the following procedure.
1) Construct a quasar template spectrum library by randomly selecting 400 quasars from a sample of SDSS
non-BAL quasars with redshifts similar to the BAL quasar candidate.
2) Use each template spectrum multiplied by SMC extinction law with a free parameter E(B-V) to model the unabsorbed spectral regions of the given candidate spectrum. The parameter E(B-V) is determined by minimizing the $\chi^2$.
3) Rank the 400 model spectra in the increasing order of $\chi^2$ and select the model spectrum, which
has the minimum difference in the BEL regions from the candidate's spectrum, among the first 20 model spectra as our final model spectrum.

Our method can choose an model spectrum which well matches the unabsorbed continuum and BELs of the BAL quasar candidate (Figure \ref{fig1}). It is worthwhile noting that the SMC curve is used to adjust the template
spectrum to match the given spectrum, so the E(B-V) parameter can be either positive or negative,
dependent on the continuum slopes of the two spectra.

The spectra of the BAL quasar candidates are then normalized by the best model spectra. In the literature, various selection criteria were
adopted to identify BALs in the normalized spectra.
We use the modified ``balnicity index" ($BI$; Weymann et al. 1991) for \ion{C}{4} BALs
and modified  ``absorption index"($AI$; Zhang et al. 2010) for \ion{Mg}{2} BALs.
Only \ion{C}{4} absorption which dips below 10\% of the
model spectrum and is at least 2000 km/s broad located between 0 and 30,000
km/s blueward of the quasar redshift is included in the $BI$ calculation, while \ion{Mg}{2} absorption at least
1600 km/s broad located between 0 and 20,000 km/s is included in AI calculation.
We use different selection criteria to identify \ion{C}{4} and \ion{Mg}{2} BALs, because it is well known that
\ion{Mg}{2} BALs are generally weaker and narrower than \ion{C}{4} BALs (Voit et al. 1993; Trump et al. 2006; Gibson et al. 2009, hereafter G09; Allen et al. 2011, hereafter A11). In the $BI$ calculation, the integration starts from zero velocity (see also Trump et al. 2006; G09; DiPompeo et al. 2012) rather than the traditional velocity of 3000 km/s (Weymann et al. 1991). Such choice can lead to a higher completeness of low velocity BAL quasars compared to previous works (see Appendix B). Additionally, according to the comprehensive study in Zhang et al. (2010), using a condition of continuous absorption width of 1600 km/s in $AI$ calculation leads to a good trade-off between the correctness and completeness of LoBALs.

Quasars with $BI>0$ and $AI=0$ are classified as HiBAL quasars, and quasars with both larger than zero are classified as LoBAL quasars. Finally, our working sample contains 2099 BAL quasars, 264 of which are LoBAL quasars. Accordingly, 16352 quasars with $BI=0$ in our DR7-WISE quasar catalog are classified into the non-BAL quasar sample. In Figure \ref{fig1}, we show the spectra of six BAL quasar candidates, four of which are finally classified as BAL quasars by our algorithm. We overplot the chosen model spectra for comparison. And the spectral regions masked out in the fitting process are shown in gray (see Appendix A for details). The chosen model spectra look quite similar to the observed spectra. They illustrate that our algorithm can effectively detect BALs with various profiles and velocities, e.g. multiple troughs BAL: SDSS J011605.94+133402.3, high velocity BAL: SDSS J082841.53+451233.9 and BAL superimposed to BEL: SDSS J010727.38-105743.1, and reject quasars with only narrow absorption, e.g. SDSS J004613.47+002358.1 and SDSS J022203.23-091531.3. The quasar J035310+094418.7 has quite low signal to noise ratio, so the match is little worse.

We also measure the velocity parameters of \ion{C}{4} BALs, i.e., the minimum, maximum and absorption-depth-weighted average blueshift velocities ($V_{\rm min}$, $V_{\rm max}$, and $V_{\rm ave}$). The positive velocities indicate blueshifted absorption with respect to the quasar's systemic redshift.
In order to further examine the reliability of our method and measurement, we compare our working
sample with other samples in Appendix B in great detail. The comparison shows that our method is at least as good as those employed in previous works. We also estimate the uncertainties of the measured BAL parameters using 500 simulated BAL quasar spectra. The simulated spectra are
constructed via multiplying the randomly selected non-BAL quasar spectra by broad absorption profiles. To mimic BALs in a realistic way, the absorption profiles are extracted from randomly selected BAL quasars. The simulation shows that the typical $1\sigma$ errors for
the maximum, minimum velocities and the balnicity are 3.6\%, 10.9\% and 6.7\% respectively.
Please see Appendix C for how we exactly perform the simulation and calculate the uncertainties.

\subsection{Measurements of UV and Infrared Emission}

We are interested in the correlation between outflow and NIR emission. All of BAL quasars in our sample have reliable magnitude measurements at
three WISE infrared bands, $W1$, $W2$ and $W3$. The redshifts of our quasars are close to 2, so the rest-frame wavelengths of the three bands all move into the NIR band, covering the wavelength region from 1$\mu$m to 4$\mu$m. We convert the magnitudes to monochromatic luminosities ($L_{\lambda}$, where $\lambda= W1$, $W2$ and $W3$), and then use a power law ($L_{\lambda}\propto\lambda^{\beta_{\rm NIR}}$)
to fit the NIR SED in the rest-frame. The NIR spectral slope, $\beta_{\rm NIR}$, is thus obtained.
As discussed in Paper I, hot dust emission begins to become dominant at wavelength larger than 1$\mu$m, compared to
the big blue bump (Elvis et al. 1994), and host galaxy contribution is negligible for $z\sim2$ quasars (Hao et al. 2014).
Therefore, the NIR spectral slope is a good indicator of the amount of hot dust emission relative to the amount of emission
from accretion disk, in the sense that $\beta_{\rm NIR}$ increases with increasing hot dust emission.
In Paper I, for comparison, we also introduced an infrared to UV luminosity ratio, CF3$=log(L_{\rm W3}/L_{\rm 2500})$,
to indicate the hot dust emission relative to the disk emission.
We found that $\beta_{\rm NIR}$ shows stronger correlations
with outflow properties than CF3, and suggested that $\beta_{\rm NIR}$ is less affected by variation, inclination effect and dust extinction (see the details in Section 3 of Paper I).

The other interested parameters are the UV continuum shape and quasar fundamental parameters, such as black hole mass and Eddington ratio. Quasar UV continua are often approximated by power laws.
Here, we use the power law, $\lambda F_{\lambda} \propto \lambda^{\beta_{\rm UV}}$, to fit the continua in two windows, [1690, 1700]\AA~ and [2225, 2250]\AA. The index $\beta_{\rm UV}$ can be used to characterize the continuum shape.
The two continuum windows are chosen to avoid strong BELs, \ion{Fe}{2} multiplets and BALs.
We use the empirical formula, presented in Wang et al. (2009a), to compute black hole masses ($M_{\rm BH}$).
The full width at half maximum (FWHM) of \ion{Mg}{2} BELs and the monochromatic luminosity at 3000\AA~($L_{\rm 3000}$), used for $M_{\rm BH}$ estimation, are taken from SH11. Eddington ratio is evaluated as $l_{\rm E} = L_{\rm BOL}/L_{\rm Edd}$, where $L_{\rm Edd}$ is Eddington luminosity
and the bolometric luminosity $L_{\rm BOL} = 5.9 \times L_{\rm 3000}$ (McLure \& Dunlop 2004).
Since the \ion{Mg}{2} line profiles of LoBAL quasars can not be well determined, we estimate these parameters only for HiBAL quasars.
We also estimate $\beta_{\rm NIR}$, $\beta_{\rm UV}$, $M_{\rm BH}$ and $l_{\rm E}$ for our non-BAL quasar sample.

\begin{deluxetable*}{l ccccc}
\tabletypesize{\scriptsize}
\tablecaption{Correlation coefficients between $l_{\rm E}$ and outflow parameters
\label{tab1} }
\tablewidth{0pt}
\startdata
\hline
Sample  &$V_{\rm max}$ & $V_{\rm min}$ & $V_{\rm ave}$ &$BI$ & \\
\hline
HiBAL   & 0.20(3.1E-17) & 0.15(4.1E-10) &  0.19(1.8E-16) & 0.10(1.9E-05)& \\
\cutinhead{Correlation coefficients between $\beta_{\rm UV}$ and outflow parameters}
Sample  &$V_{\rm max}$ & $V_{\rm min}$ & $V_{\rm ave}$ &$BI$& \\
\hline
Whole   & -0.09(3.4E-05) & -0.23(3.5E-26) &  -0.15(1.5E-12) & 0.10(1.2E-05) &\\
HiBAL   & -0.14(1.4E-07) & -0.21(4.0E-19) &  -0.17(1.5E-13) & 0.03(1.4E-02) &\\
LoBAL   & -0.08(3.8E-01) & -0.27(2.4E-03) &  -0.04(6.7E-01) & 0.01(4.6E-04) &\\
\cutinhead{Correlation coefficients between $L_{\rm 3000}$ and outflow parameters}
Sample  &$V_{\rm max}$ & $V_{\rm min}$ & $V_{\rm ave}$ &$BI$ &\\
\hline
Whole   & 0.13(4.5E-08) & 0.05(3.1E-02) & 0.10(4.3E-06) & -0.03(1.4E-01) & \\
HiBAL   & 0.13(2.8E-08) & 0.08(1.0E-03) & 0.11(2.1E-06) & -0.06(7.5E-03) & \\
LoBAL   & 0.06(3.2E-01) & -0.10(8.2E-02) & 0.01(8.4E-01) & -0.04(5.1E-01)&\\
\cutinhead{Correlation coefficients between $M_{\rm BH}$ and outflow parameters}
Sample  &$V_{\rm max}$ & $V_{\rm min}$ & $V_{\rm ave}$ &$BI$ & \\
\hline
HiBAL   & -0.09(8.2E-05) & -0.09(4.5E-04) &  -0.10(2.6E-05) &  0.15(3.0E-10)& \\
\cutinhead{Correlation coefficients between $\beta_{\rm NIR}$ and outflow parameters}
Sample &$V_{\rm max}$ & $V_{\rm min}$ & $V_{\rm ave}$ &$BI$& \\
\hline
Whole  &   0.19(4.3E-19) & 0.21(3.5E-23) & 0.22(3.6E-24) & 0.14(4.4E-10)&\\
HiBAL  &   0.20(1.1E-17) & 0.19(1.5E-16) & 0.21(5.7E-20) & 0.15(2.1E-10)&\\
LoBAL  & 0.19(1.9E-03) & 0.33(4.4E-08) & 0.28(4.8E-06) & 0.21(4.6E-04)& \\
\hline
\cutinhead{Correlation coefficients between $\beta_{\rm NIR}$ and other quasar physical parameters}
Sample &$\beta_{\rm UV}$&$l_{\rm E}$ & $M_{\rm BH}$ & $L_{3000}$ \\
\hline
Whole  & -0.17(2.0E-13) & ---         &    ---         & -0.13(8.4E-09) \\
HiBAL  & -0.17(4.5E-13) & -0.04(6.7E-02) & -0.07(2.0E-03) & -0.11(6.5E-09) \\
LoBAL  & -0.27(3.9E-05) & ---         &   ---          & -0.22(2.4E-04)
\enddata
\tablenotetext{Note.} {For each entry, we list the Spearman rank correlation coefficient ($\rho_{\rm r}$) and
the probability of the null hypothesis that the correlation is not present ($P_{\rm null}$).}
\end{deluxetable*}

\section{Correlation Analysis}
\subsection{Properties of BAL quasars}

\figurenum{2}
\begin{figure}[tbp]
\epsscale{1.} \plotone{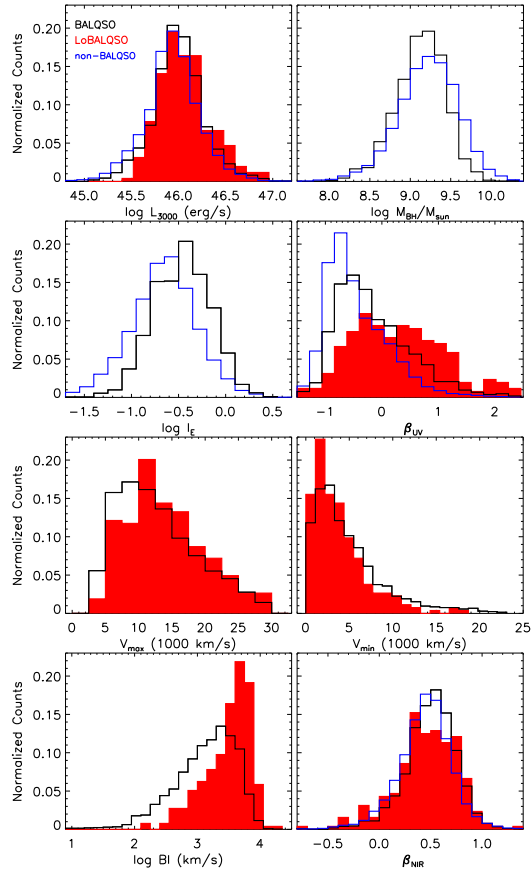}
\caption{The probability distributions of the quasar fundamental properties,
the continuum slopes and the outflow parameters, as indicated in the panels.
The black and red curves represent our whole BAL quasar sample and LoBAL quasar subsample, respectively.
In the $M_{\rm BH}$ and $l_{\rm E}$ panels, we only show the distributions for HiBAL quasars.
For comparison, the distributions for non-BAL quasars are overplotted in blue dashed lines.
}\label{fig2}
\end{figure}

In Figure \ref{fig2}, we show the probability distributions of the measured parameters for both BAL and non-BAL quasars in our DR7-WISE quasar catalog, if available. The blue, black and red histograms represent non-BAL quasars, BAL quasars and LoBAL quasars, respectively. Generally, there are systematic differences in the shown properties between BAL and non-BAL quasars. BAL quasars, on average, have higher luminosities and Eddington ratios than non-BAL quasars, consistent with previous works (e.g., Reichard et al. 2003; Trump et al. 2006; Gangly et al. 2007;
G09; A11). These results favor a radiatively driven outflow scenario. Meanwhile, an opposite and weaker trend is found in the black hole mass parameter. Because the SDSS spectroscopical survey is magnitude limited, low $l_{\rm E}$ quasars tend to
have high $M_{\rm BH}$ (see e.g. Dong et al. 2009). Thus the dependence on $M_{\rm BH}$ shown in this work might be not intrinsic but a combined effect of the dependence on $l_{\rm E}$ and the false correlation between $M_{\rm BH}$ and $l_{\rm E}$. However Gangly et al. (2007) found a different mass dependence. We note that the mass dependence is not detected in the SDSS DR5 low-$z$ LoBAL quasar sample (Zhang et al. 2010).

The $BI$ parameters are estimated in the same way as $BI_0$ used in G09, with the only difference being in integration range. We calculate $BI$ in the velocity range from 0 km/s to 30,000 km/s, while G09 adopted a smaller range from 0 km/s to 25,000 km/s. Since outflows with velocities larger than 25,000 km/s are relatively rare, this difference would not significantly
impact the parameter measurements as long as we have the same model spectra as theirs. It is thus useful to compare our parameters with theirs. Our $BI$ distribution peaks at about 3000 km/s and G09's distribution peaks at the velocity range from 2000 to 4000 km/s. Meanwhile, our $V_{min}$ distribution peaks at about 2000 km/s, also consistent with G09. However, the $V_{max}$ distribution looks different from G09's. Our distribution has an apparent peak at $\sim$9000 km/s, while G09 presented an almost uniform distribution in the velocity range from 3000 km/s to 25,000 km/s. The difference may be partly due to the fact that our method can more effectively detect the low-velocity BALs.
In Appendix B, we present a much more detailed comparison of our measurements and the other measurements available in the literature (G09 and A11).

Consistent with previous works (Weymann et al. 1991; Reichard et al. 2003; Trump et al. 2006; G09; Zhang et al. 2010; Baskin et al. 2013), our results show that BAL quasars are, on average, redder in UV spectral region than non-BAL quasars. It suggests the existence of a significant amount of obscuring dust associated with BAL (especially LoBAL) outflows.
Besides the UV continuum slope, LoBAL quasars have different $BI$ distribution from the entire BAL quasar population.
LoBAL quasars are more frequently detected in BAL quasars with stronger \ion{C}{4} BALs (see also G09 and Baskin et al. 2013). The mean $V_{max}$ of \ion{C}{4} BALs for LoBAL quasars is 13,136 km/s,  higher than that for the entire sample, 11,785 km/s. And the mean $V_{min}$ for LoBAL quasars is 2903 km/s, lower than that the mean value of the entire sample, 3927 km/s (see A11 for a similar result).
The differences in velocities and absorption strength imply that LoBAL outflows are stronger and of larger column density than HiBAL outflows.

The distributions of the NIR slopes, $\beta_{\rm NIR}$, appear to be similar between non-BAL and BAL quasars (the bottom right panel of Figure \ref{fig2}). The mean $\beta_{\rm NIR}$ for these two samples are 0.43 and 0.49, and the standard deviations are 0.24 and 0.23. Hot dust emission in BAL quasars is only slightly stronger than that in non-BAL quasars. It is consistent with the prediction of the BAL unifying model, in which the only difference between BAL and non-BAL quasars is viewing angle (e.g. Weymann et al. 1991). The mean $\beta_{\rm NIR}$ for LoBAL is 0.45, similar to the other two samples, while the standard deviation is apparently larger, about 0.32. If BAL outflows are really observable along some specific directions, such as equatorial direction as suggested by previous works, the similar distributions further imply that the dependence of $\beta_{\rm NIR}$ on inclination angle is quite weak, even absent. More recently, Runnoe et al. (2013) investigated the inclination dependence of quasar SED in a small radio-loud sample. Using the radio core dominance to indicate the orientation, they show in their figure 6 that the NIR (1 $\sim$ 4$\mu$m) slope of the face-on quasars is slightly less than that of the edge-on quasars, broadly consistent with the expectation from our result.

\subsection{Correlations of outflow with quasar physical properties}

\figurenum{3}
\begin{figure*}[tbp]
\epsscale{1.} \plotone{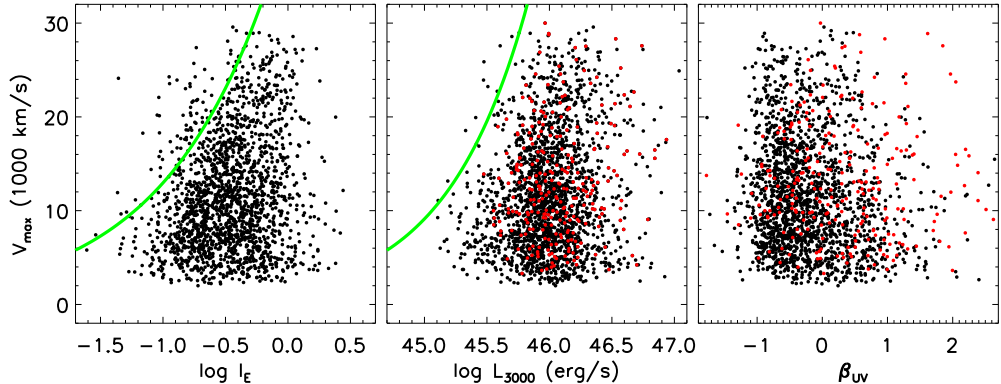}
\caption{Correlations of the maximum velocities of BALs with
the Eddington ratios (left), the luminosities at 3000\AA~(middle).
the UV spectral slopes (right),
Black and red dots represent HiBAL quasars and LoBAL quasars respectively.
The green curves show the predicted scaling relationship between $V_{\rm max}-l_{\rm E}$ from Hamann (1998), Misawa et al. (2007) (middle)
and $V_{\rm max}-L_{\rm 3000}$ fitting curve from Laor \& Brandt (2002) (right). The normalization of the curves are taken from Gangly et al. (2007).}
\label{fig3}
\end{figure*}

Lots of observational works have been devoted to understanding the launch and acceleration processes of quasar outflows.
The correlations of outflow properties with Eddington ratio, luminosity and SED shape have been reported in the literature
(Laor \& Brandt 2002; Gangly et al. 2007; Fan et al. 2009; Zhang et al. 2010; Baskin et al. 2013). These findings are largely
confirmed by our new BAL quasar sample. In Table \ref{tab1}, we list the Spearman correlation coefficients and the probabilities
of the null hypothesis for correlations between the BAL parameters ($V_{\rm max}$, $V_{\rm min}$,
$V_{\rm ave}$ and $BI$) and quasar properties ($l_{\rm E}$, $L_{\rm 3000}$,
and $\beta_{\rm UV}$). One can find moderate or weak (but significant) correlations between these two sets of parameters.
A quasar with higher $l_{\rm E}$ or higher $L_{\rm 3000}$ or bluer UV spectrum has a tendency to harbor a stronger outflow.
Generally, the velocity parameters are more strongly correlated with the quasar properties than $BI$. Moreover, the correlation with
$l_{\rm E}$ is, on average, stronger than the correlations with $L_{\rm 3000}$ and $\beta_{\rm UV}$.
And $M_{\rm BH}$ is almost uncorrelated with the velocity parameters and weakly related to $BI$. All of these results are broadly
consistent with the findings in Ganguly et al. (2007), although our correlation coefficients are slightly smaller.
For further comparison, we show $V_{\rm max}$ against $l_{\rm E}$, $L_{\rm 3000}$ and $\beta_{\rm UV}$ in Figure \ref{fig3}.
Following Ganguly et al., we also plot the predicted scaling relation between $V_{\rm max}$ and  $l_{\rm E}$ (the normalization is taken from Ganguly's figure 6) and the $V_{\rm max}$-$L_{\rm 3000}$ fitting curve (see their figure 7 and equation 5) in the corresponding panels. These two curves appear to be the upper envelopes of these scatter points, again consistent with Ganguly et al..

\figurenum{4}
\begin{figure}[tbp]
\epsscale{1.} \plotone{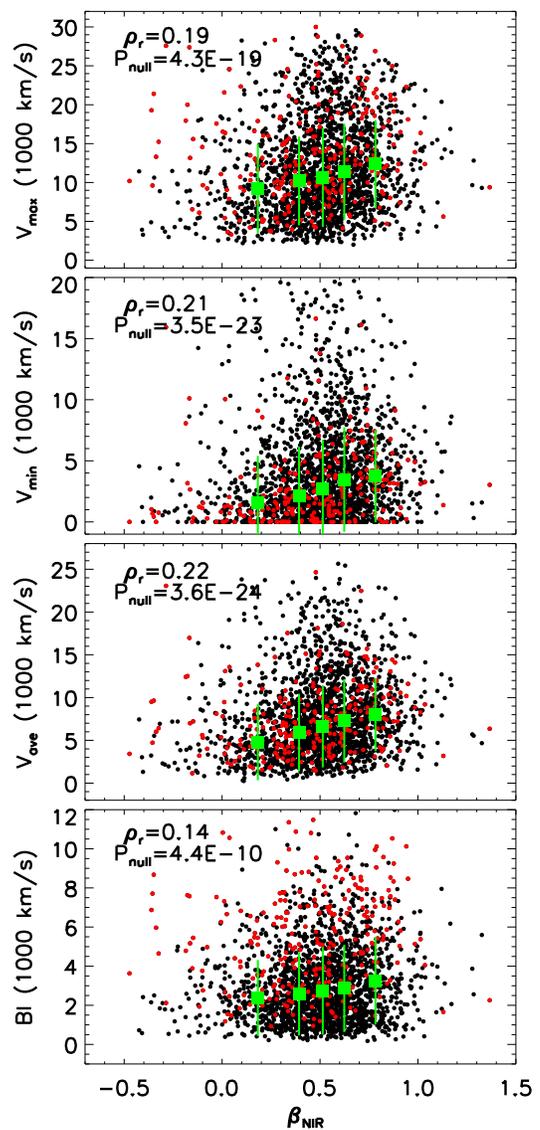}
\caption{Relationship between $\beta_{\rm NIR}$ and outflow properties.
Black and red dots represent HiBAL quasars and LoBAL quasars respectively.
The green squares show the median values and the error bars indicate the standard deviations.
}\label{fig4}
\end{figure}

Now we explore the role of $\beta_{\rm NIR}$ in the nature of outflows.
Figure \ref{fig4} shows the four BAL parameters against the NIR spectral slope, $\beta_{\rm NIR}$.
HiBAL and LoBAL quasars are shown in black and red dots respectively. We divide the whole BAL quasar sample into five equal-sized
subsamples according to $\beta_{\rm NIR}$, then calculate the median values of $\beta_{\rm NIR}$ and BAL parameters. The results are shown in filled green square (Figure \ref{fig4}). The error bars indicate the standard deviations of the corresponding parameters.
Despite of large scatter, clear trends between these BAL parameters and $\beta_{\rm NIR}$ are present.
One can find that a quasar with stronger hot dust emission relative to disk emission (i.e. larger $\beta_{\rm NIR}$) has a tendency to harbor a stronger outflow, which has higher maximum, minimum and average velocities and is able to yield stronger absorption. As discussed earlier, we found a strong correlation between $\beta_{\rm NIR}$ and blueshift and asymmetry index (BAI) of \civ\ BELs in non-BAL quasars, especially in high Eddington ratio quasars (see paper I). If the blueshifted \civ\ BELs are indeed emitted by the same outflow components as those producing \civ\ BALs (Wang et al. 2011),  one would readily expect the correlations shown in this work. Our results thus confirm the connection between outflow strength and hot dust emission from a very different aspect.

We use the Spearman rank correlation tests to evaluate the correlation strength (see Figure \ref{fig4} and Table \ref{tab1}). As expected, these correlations are moderate and significant. We also list the correlation coefficients for HiBAL and LoBAL quasar subsamples separately in Table \ref{tab1}. One interesting phenomenon is that the correlations for LoBAL quasars are stronger than those for HiBAL quasars. It hints that dust indeed plays an unusual role in these correlations. Although the physical meaning of BAL parameters is more specific than BAI (the outflow indicator measured from \civ\ BELs), BAL parameters are more weakly correlated with $\beta_{\rm NIR}$. Note that the coefficient for the BAI-$\beta_{\rm NIR}$ relationship is 0.55 for high Eddington ratio quasars (paper I). One possible reason is that BAL parameters are absorption properties while BAI and $\beta_{\rm NIR}$ are both emission properties. Absorption is produced by the gas along the line of sight (LOS) so the absorption properties are only dependent on the LOS structures. In contrast, emission comes from the entire emitting gas/dust, so the emission properties are able to reflect the overall properties. The different origins in the absorption and emission properties may cause large scatter and weaken the observed correlations. The second possible reason is that outflow strengths might be affected by various physical processes. It is already known that outflow strength is significantly correlated with other physical parameters as shown in previous works and this work, and $\beta_{\rm NIR}$ is almost independent of these factors (see below). In addition, the inclination dependence (see Section 4.1) and the uncertainty in BAL parameter measurements may also introduce scatter into the correlations .

Are the correlations between $\beta_{\rm NIR}$ and BAL properties intrinsic or a secondary effect induced by other relationships?  For instance, if $\beta_{\rm NIR}$ tightly correlates with
$l_{\rm E}$, the correlations of BAL properties with $\beta_{\rm NIR}$ could be interpreted as being induced by the correlations with $l_{\rm E}$. It is therefore necessary to check whether $\beta_{\rm NIR}$ correlates with other quasar physical properties that are found to be related to BAL properties, such as $l_{\rm E}$, $L_{\rm 3000}$ and $\beta_{\rm UV}$. We list the corresponding correlation coefficients in Table \ref{tab1}. As one can see that the correlation between $\beta_{\rm NIR}$ and $l_{\rm E}$ is absent, consistent with that in non-BAL quasars (paper I). And $\beta_{\rm NIR}$ only moderately or weakly correlates with $\beta_{\rm NIR}$ and $L_{\rm 3000}$. Therefore the secondary effect hypothesis is unlikely. More importantly, in most cases, the correlations between BAL properties and $\beta_{\rm NIR}$ are almost the strongest among all the correlations that we investigate (see Table \ref{tab1}). It also favors that the correlations of $\beta_{\rm NIR}$ with outflow properties is intrinsic and can not be induced by the other correlations shown in this paper.

The anti-correlation between $\beta_{\rm NIR}$ and $\beta_{\rm UV}$ itself is interesting. In a dust extinction and re-emission scenario, a reasonable expectation is a positive correlation, which however is opposite to what we find. It means that the observed correlation is possibly ascribed to another effect. For a given UV luminosity, the disk emission of a blue quasar is weaker at NIR band than that of a red quasar. Since $\beta_{\rm NIR}$ reflects the ratio of hot dust emission to disk emission at NIR band, $\beta_{\rm NIR}$ tend to be larger in the blue quasar. The observed correlation is naturally yielded and does not conflict with the dust extinction and re-emission scenario. The negative trend therefore suggests that dust extinction only has little impact on the UV SED, on average. In fact, the difference in $\beta_{\rm UV}$ distributions between non-BAL and BAL quasars is much smaller than the distribution range of $\beta_{\rm UV}$ (the right panel of the second row in Figure \ref{fig2}). It is consistent with previous result that the average amount of reddening in HiBAL quasars is only $E(B-V)\sim$0.023 (Reichard et al. 2003). All of these suggest that the UV slope is not determined by dust extinction but other factors, such as accretion process. We further check the trend and find that the correlation is largely dominated by a small fraction of quasars with small $\beta_{\rm NIR}$. After excluding quasars with $\beta_{\rm NIR}<0.1$ (about 6\% of BAL quasars), the correlation coefficient for
$\beta_{\rm NIR}$-$\beta_{\rm UV}$ relationship is reduced to about $0.1$.

\subsection{Composite spectra}

\figurenum{5}
\begin{figure*}[]
\epsscale{1.} \plotone{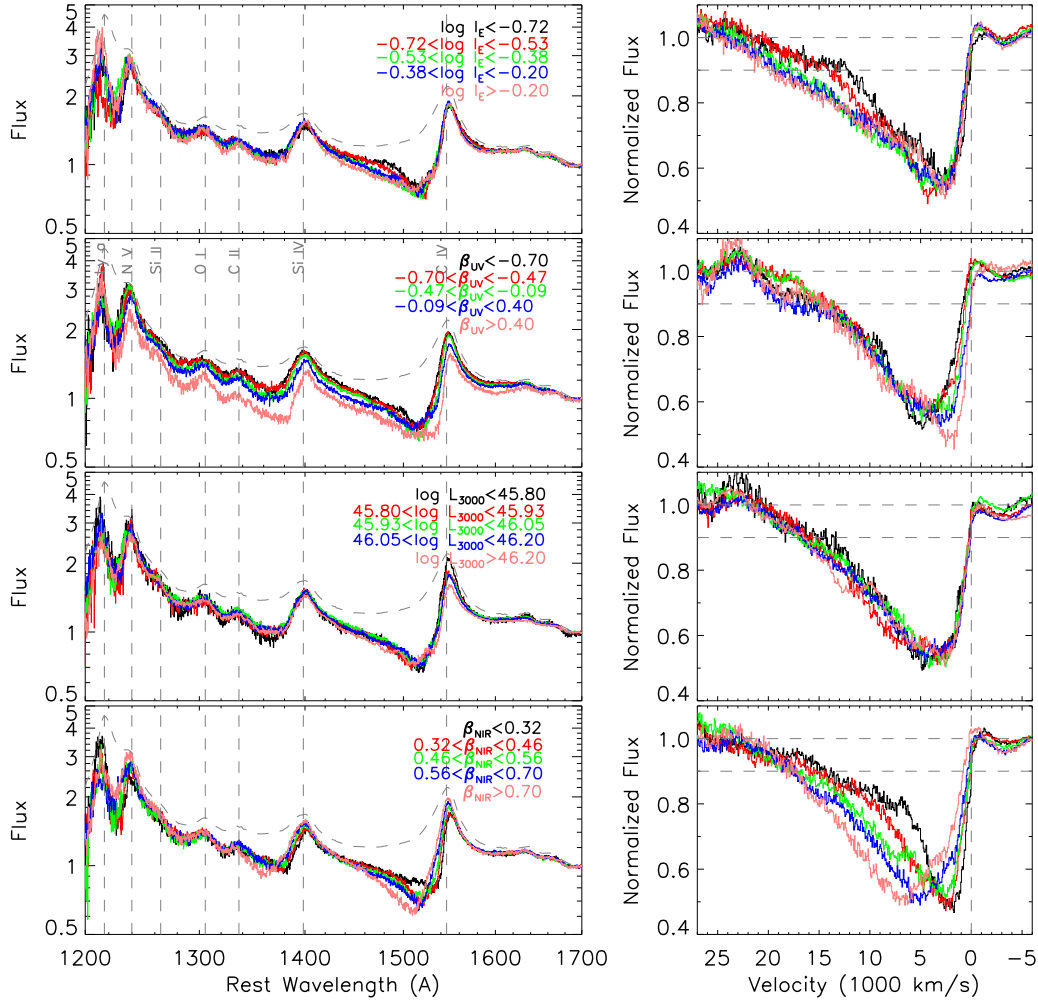}
\caption{Composite spectra of BAL quasar subsampes according to Eddington ratio, the UV slope, 
the luminosity at 3000\AA~ and the NIR slope, and their normalized flux of \ion{C}{4} BALs. Please see the text on how we construct the composites. For comparison, the SDSS DR7 composite spectrum is shown by gray dashed curve in left panels,
we also show the lines of the velocity $V=0$ km/s, the normalized flux $f_N=1$ and $f_N=0.9$
in right panels.
}\label{fig5}
\end{figure*}

As shown in the above subsection, we present the correlations of outflow properties with quasar physical properties.
There is large scatter in these correlations and the correlation strength is only moderate or weak (however significant).
One possible reason for this is that the outflow properties measured from the absorption troughs are sensitive to the LOS structures.
If outflows are clumpy as suggested by previous works (e.g. Smith \& Raine 1988; Netzer 1993; Krolik \& Kriss 1995; Netzer 1996; Elvis 2000; Chelouche \& Netzer 2005),
the BAL parameters may significantly vary with the viewing angle.
In other works, for a given quasar outflow, the BAL parameters may be very different along different LOS. One useful method to suppress this
LOS effect is to construct composite spectra. Composite spectra may eliminate the variation in outflow properties and distinctly tell us how the mean absorption profile at different velocity changes with different physical parameters.

\begin{figure*}[tbp]
\epsscale{1.} \plotone{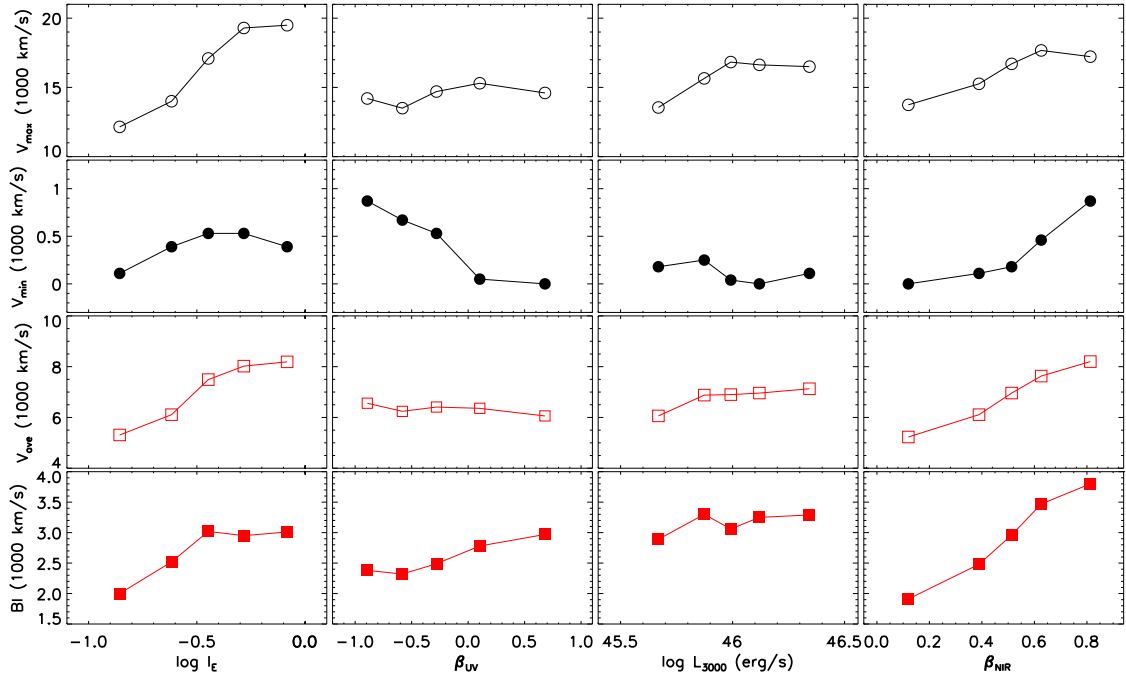}
\caption{Outflow parameters measured from composite spectra (Figure \ref{fig5})
as functions of quasar physical properties.
}\label{fig6}
\end{figure*}

The whole BAL quasar sample is divided into five equal-sized subsamples, according to the UV slope,
Eddington ratio (here, only HiBAL quasars are used), the luminosity at 3000\AA~ and the NIR slope, respectively. We adopt the method of Vanden Berk et al. (2001) to create the geometric mean composite spectra for each subsample.
For each BAL quasar, we deredshift its observed spectrum into its rest frame. The spectrum is then normalized at 1700\AA, rebinned into the same wavelength grids, and geometrically averaged bin by bin. In the left panels of Figure \ref{fig5}, we plot composite spectra of these subsamples in sequence. The SDSS DR7 composite spectrum is overplotted in gray dashed line for comparison.
We then use the exactly same method as shown in Section 2.1 and Appendix A to obtain the model spectrum for each composite spectrum. The normalized composite spectra around \civ\ BALs are shown in the right panels of Figure \ref{fig5} and the measured BAL parameters ($V_{\rm max}$, $V_{\rm min}$, $V_{\rm ave}$ and $BI$) as functions of $\beta_{\rm UV}$, $l_{\rm E}$, $L_{\rm 3000}$ and $\beta_{\rm NIR}$ are shown in Figure \ref{fig6}.

One can see, from Figure \ref{fig5}, that the dependence of the mean absorption profile on $\beta_{\rm NIR}$ is the strongest. The absorption shifts blueward as a whole with increasing $\beta_{\rm NIR}$. Both the maximum and minimum velocities significantly increase with the increase of $\beta_{\rm NIR}$. And the absorption profile apparently broadens as $\beta_{\rm NIR}$ increases. We also check \siiv\ BALs and find the same trend. In particular, the trend for \siiv\ is almost as strong as that for \civ.
The second most important factor is Eddington ratio. Eddington ratio seems to have a dramatic impact on the large velocity part of outflows.
The maximum velocity increases with increasing Eddington ratio.
Interestingly, the absorption profile almost remains unchanged at velocity less than 10,000 km/s as $l_{\rm E}$ varies. On the contrary, $\beta_{\rm UV}$ only affects the low velocity part of the outflowing gases. The absorption trough in a red quasar is closer to zero velocity than that in a blue one. And the absorption profile is almost independent of $L_{\rm 3000}$. The dependencies of the mean absorption profile on quasar physical parameters may offer valuable insight into the launch and acceleration of outflows.

The impression obtained from the composites is confirmed by the quantitative results shown in Figure \ref{fig6}.
Outflow velocities and strength measured from the composites increase with the enhancement of black hole accretion and hot dust emission.
In particular, $\beta_{\rm NIR}$ is the factor which has the strongest impact on $BI$. The parameter $BI$ for the largest $\beta_{\rm NIR}$ composite is more than twice as large as that for the smallest $\beta_{\rm NIR}$ composite. The maximum velocity is more sensitive to $l_{\rm E}$ than other factors. $V_{\rm max}$ increases from $\sim$12,000 km/s to $\sim$20,000 km/s. And $\beta_{\rm NIR}$ and $l_{\rm E}$ appear to be both dominant for $V_{\rm ave}$. In the case of $V_{\rm min}$, $\beta_{\rm UV}$ becomes as important as $\beta_{\rm NIR}$. One can see that $V_{\rm min}$ changes from 0 km/s to nearly 900 km/s with decreasing $\beta_{\rm UV}$ and increasing $\beta_{\rm NIR}$. Overall, the dependence of outflow properties on $\beta_{\rm NIR}$ is more significant.
We note that these results are well consistent with the statistical results shown in the above subsection (see Table \ref{tab1}).

\figurenum{6}

Recently, Baskin et al. (2013) also analyzed the average absorption profiles of DR7 BAL quasars. From their figure 9 and 11, one can clearly see that the UV slope (measured in the 1700-3000\AA~range) only impacts the small velocity part while the Eddington ratio primarily controls the large velocity part. Additionally, they found no significant dependence on the luminosity. These trends are well consistent with what we show here. However their results also have apparent difference from ours, in particular in the behavior associated with the UV slope. They found that the low velocity part of the absorption deepens as the UV continuum becomes redder. We do not detect this trend in our composite. The difference may be due to several reasons. First, we use very different method to construct the composite.
Second, we use very different method to derive the normalized spectra. And third, about 17\%
BAL quasars in SH11 sample (used by Baskin et al.) are excluded by our algorithm and additional 117 quasars are added in(see Table A1).

Another interesting result found by Baskin et al. is the significant dependence of absorption profile on the equivalent width (EW) of \heii\ BEL, which is interpreted in terms of the dependence on ionization SED. It will be interesting to check whether there is any correlation between EW(\heii) and $\beta_{\rm NIR}$. \heii\ BELs are blended with \oiii\ and strong \civ\ BELs, so it is hard to properly decompose
\heii\ BEL. We therefore use the integrated flux between 1620\A\ and 1650\A\ to calculate EW(\heii), which is used in Baskin et al. (Please see their description of the method for details). The EW(\heii) measured from the composites of different $\beta_{\rm NIR}$ are
5.15, 5.56, 5.35, 5.91 and 5.55\A, in sequence of increasing $\beta_{\rm NIR}$.
The NIR slope is apparently uncorrelated with EW(\heii). To further verify the result, we also measure EW(\heii) for all individual BAL quasars. The Spearman correlation coefficient for these two parameters is only -0.01 (the null
probability is 0.78).

\section{Discussion}

In Paper I, we adopted the blueshift and asymmetry index ($BAI$) of \ion{C}{4} BELs to indicate the outflow property and velocity, and
found an interesting correlation between BAI and $\beta_{\rm NIR}$. In this paper, the blueshifted velocities and absorption strength of BALs instead of BAI are used to characterize the outflow strength. Again, we find significant correlations with $\beta_{\rm NIR}$, well consistent with the results shown in Paper I. Using outflow parameters measured from BELs (e.g. BAI) and BALs (e.g. $V_{\rm max}$, $V_{\rm ave}$ and $BI$) both have advantages and shortcomings. On one hand, BELs are the integral of emission over entire volume of the outflow and represents the overall properties, while BAL troughs only hold the information of outflow along the LOS and thus are sensitive to clumpy structures. It may be the reason for that $\beta_{\rm NIR}$ is more strongly correlated with BEL parameters than BAL parameters. On the other hand, the physical meanings of BAL parameters are more straightforward and specific than those of BEL parameters.
The emission from virialized gas also contributes to BELs so that it is generally hard to reliably extract outflow parameters from BELs. In contrast, the contamination in BALs from other line features is usually unimportant and easily to handle (see Section 2). It is one of the reasons that we think it is essential to use BAL quasars to revisit the relationship between outflow and hot dust.
Overall, these two works together present robust evidences for the important correlation and connection between outflow strength and hot dust emission.

In order to understand the origin of these correlations, we present detailed discussions in the following. In Section \ref{sec_df}, we examine whether or not a dust-free outflow, together with a hydrostatic torus model, can properly accommodate our findings. Our analysis based on current observational results suggest that this scenario is likely unviable. In Section \ref{sec_do}, we discuss the dusty outflow scenarios, which have ever been proposed in Paper I.

\subsection{Dust-free outflow scenario}\label{sec_df}

Suppose that hot dust emission in these high-redshift quasars has nothing to do with outflows, and is predominantly emitted by the innermost part of a hydrostatic and optically thick torus, as suggested by lots of previous studies (e.g., Neugebauer et al. 1987; Barvainis 1987; Suganuma et al. 2006; Kishimoto et al. 2007; Mor et al. 2009; Mor \& Trakhtenbrot 2011). It would be interesting to see whether or not the correlations shown in this paper and Paper I are induced by a third factor that simultaneously governs or relates to outflows and dust emission.

We first discuss the inclination effect. The NIR slope, $\beta_{\rm NIR}$, measures the amount of hot dust emission relative to accretion disk emission. The disk emission is theoretically predicted to scale with the cosine of the disk inclination angle (IA). If hot dust emission is (approximately) isotropic, $\beta_{\rm NIR}$ is expected to increase with increasing IA. However, dust torus may block the hot dust emission at a large IA (Roseboom et al. 2013). It complicates the situation and makes the dependence of $\beta_{\rm NIR}$ on IA uncertain. More recently, Runnoe et al. (2013) used a sample of radio-loud quasars to investigate the inclination dependence of quasar spectral energy distribution. The composites in their figure 6 clearly present a weak dependence of $\beta_{\rm NIR}$ on IA, in the sense that a edge-on quasar spectrum tend to show a slightly larger $\beta_{\rm NIR}$ than a face-on one. We thus still assume that $\beta_{\rm NIR}$ increases with increasing IA
Since the outflow covering factor is usually small, about 10\%$\sim$20\% (e.g., Tolea et al. 2002; Hewett \& Foltz 2003; Reichard et al. 2003; Trump et al. 2006), it is unlikely that outflow properties and $\beta_{\rm NIR}$ systematically and significantly vary with IA within the opening angle of outflows. However, one can still elaborately design a outflow model to match the observation. For example, the mean outflow IAs vary among quasars and the small-IA outflows are, on average, weaker than the large-IA outflows.
The problem for this hypothesis is that there is no any observational evidence for it and the inclination dependence of $\beta_{\rm NIR}$ is actually weak (Runnoe et al. 2013). In addition, it also fails to account for the correlation with BAI, the outflow parameter measured from BELs (see Paper I). The inclination effect might be the other factor for the large scatter in the observed correlations.

Lots of studies have found the correlations of outflow properties with quasar fundamental parameters
e.g., Reichard et al. 2003; Gangly et al. 2007; Fan et al. 2009; A11; Baskin et al. 2013).
Outflow strength is significantly dependent on Eddington ratio, broad band SED (e.g. ionization SED, and UV continuum slope, $\beta_{\rm UV}$) and luminosity. If these factors also have an impact on hot dust emission, one might find a piece of evidence to support the dust-free outflow scenario. As shown in this paper and previous studies (e.g. Mor \& Trakhtenbrot 2011; Paper I), both $\beta_{\rm NIR}$ and the infrared to UV flux ratio are almost independent of Eddington ratio. And the correlations between luminosity and the two hot dust indicators are either absent or negative. In addition, the correlation of outflow strength with $\beta_{\rm NIR}$ is stronger than that with $\beta_{\rm UV}$ and the correlation between the two slopes are apparently weaker (Table \ref{tab1}). Hence none of them is the factor that we are seeking. The role of ionization SED is unclear, since we do not know its relation with $\beta_{\rm NIR}$. Recently, Baskin et al. (2013) detected an interesting dependence of outflow strength on EW(\heii) and ascribed it to the impact of ionization SED. We thus check the relationship between EW(\heii) and $\beta_{\rm NIR}$. Again, no correlation is found.

Another interesting factor is metallicity. Quasars harboring strong outflows tend to have high gas metallicity (Hamann 1998; Leighly 2004; Wang et al. 2009b). Recently, Wang et al. (2012) found that \ion{C}{4} blueshift increases with gas metallicity. It means that outflows are stronger in higher metallicity environment. Since dust forms more easily in higher metallicity gas, one may expect that metallicity is
an appropriate factor which simultaneously influences outflow and hot dust emission. However, such naive expectation may not be true. For a typical torus, the relative amount of dust emission is determined by dust covering factor, but not dust amount.
Theoretically, it is unclear how to construct a relationship between dust covering factor and metallicity. And observationally, there is no evidence supporting it. On the contrary, the required positive correlation seems inconsistent with the current observational facts that metallicity strongly increases with increasing luminosity(e.g. Hamann \& Ferland 1999; Nagao et al. 2006) and dust covering factor (indicated by the infrared to UV flux ratio) is weakly anti-correlated with luminosity (e.g. Mor \& Trakhtenbrot 2011).

To sum up, the current observational results listed in this paper and the literature do not favor the dust-free outflow scenario. However further works, especially on the ionization SED and metallicity, are still required to examine our point.

\subsection{Dusty outflow scenarios}\label{sec_do}

In Paper I, we attempted to propose two plausible mechanisms to interpret the correlation of outflow strength with hot dust emission. Here we present further discussion. In the first mechanism, dust is intrinsic to outflows.
This idea is supported by the similar locations of BAL outflows and hot dust (Elvis 2000  and references therein). There is little observational constraint on the (relative) location of BAL outflows. Since the absorption material almost always absorbs the BELs, it is usually believed that BAL outflows are co-spatial with or outside of BEL regions (BLRs). The reverberation mapping results of local AGNs suggested that hot dust is also located at the boundary of BLRs (e.g., Suganuma et al. 2006).
These two implies the possibility of originally dusty outflows. Outflows may emerge from the outer region of accretion disk or even the innermost region of torus, in which the gas clouds are dusty and relatively cold. These dusty clouds are uplifted above the disk, and exposed to the central engine. The low density part is highly ionized and responsible for the blueshifted absorption and emission lines. Dust survives in the dense region and radiates in NIR band. In addition, dust perhaps forms in dense clouds embedded in outflows (Elvis et al. 2002). A strong outflow carries large amount of dust, and thus enhances the NIR emission. At the same time, the acceleration due to dust absorption and scattering make the outflow stronger, which further enhances the correlation.

It is worth noting that Elitzur \& Shlosman (2006; see also Emmering et al. 1992) presented a torus outflow model. In their model, dusty molecular clouds are uplifted from the accretion disk and injected into outflows, and these moving optically thick clouds constitute a torus. Our scenario resembles theirs. But there are apparent differences. Firstly, the velocity of the outflows we studied is of order 10000 km/s, much faster than the torus outflows. Second, the clouds in BAL outflows are unlikely to be optically thick in dust opacity, since a large fraction of BAL quasars can be detected at UV band and the average amount of reddening in BAL quasars is only $E(B-V)\sim0.023$ (Reichard et al. 2003). Thirdly, the large fraction of BAL outflows is highly ionized. One solution to the discrepancies is that the torus structure is stratified. The dusty BAL outflow is perhaps the inner edge of the torus outflow of Elitzur \& Shlosman and is in an extreme ionization and dynamical state compared to the bulk torus. In this case, it is reasonable to speculate the existence of an outflow component with ionization and velocity in between. Recently, Zhang et al. (2013) found, in low redshift AGNs, an interesting connection between mid-infrared (MIR) emission and outflows in NLR, the velocity of which is of order several hundred kilometers per second (see also H\"{o}nig \etal\ 2013).
The NLR outflows seem to fill the gap. However, a simple extrapolation from local AGNs may not be appropriate. Studies of the relationship between various outflow components (such as NLR, BAL, BEL) and MIR emission of $z\sim2$ quasars may help in uncovering the underlying physics.

Before discussing the second mechanism, let us list several observational facts. The maximum velocity of a typical BAL outflow is about 10000 km/s. The typical width of BELs is about 4000 km/s and BAL outflows are co-spatial with or outside of BEL regions. It means that the typical Keplerian velocity at BAL outflow location is less than 4000 km/s. The last fact is the absence of evidence for outflows with similar velocity at galactic scale or in NLR. All of these facts together show that BAL outflows are strongly decelerated before reaching NLR and the gravity of their central black holes, however, is not responsible for the deceleration. Thus it is likely that outflows interact with surrounding medium. This medium is located between BLR and NLR and should be dense enough that it can effectively decelerate the outflows. Apparently, interaction with torus clouds is the most favorable.

Recent hydrodynamical simulation (Wagner et al. 2013) showed that fast outflows can break dense clouds into diffuse warm filaments. This process makes more dust in the clouds exposed to the central UV source. As a consequence, the infrared emission increases and the outflows become dusty. Since a stronger outflow can ablate dense clouds more effectively, the dependence of NIR emission on outflow strength is yielded. In order to reach a high enough temperature to radiate in NIR band, the interaction has to occur at the innermost region of torus.
In this scenario, outflows confine the geometry and subtending angle of dusty torus.
It is in line with the suggestion that outflows blow away ambient dusty gas and transform a buried quasar into normal quasar phase (e.g. Sanders et al. 1988; Rupke \& Veilleux 2013). One problem of this scenario is that the interaction timescale (see e.g. Wagner et al. 2013) is much shorter than the quasar lifetime. The correlation might disappear after outflows blow away all of clouds in the outflow direction.

\section{Summary}

Based on a large z$\sim$2 BAL quasar sample built from the Wide-field Infrared Survey and the
Sloan Digital Sky Survey data Release Seven, we investigate the correlations of outflow strength on hot
dust emission (indicated by the near-infrared continuum slope, $\beta_{\rm NIR}$) and other quasar
properties, such as Eddington ratio, luminosity and UV continuum slope. We collect BAL quasars from
two existed BAL quasar samples in the literature, then we use our new automatic algorithm to refine
the sample and measure the BAL parameters. The final working sample contains 2099 BAL quasars.

We first reexamine the differences between BAL and non-BAL quasars and the dependencies of outflow
properties on Eddington ratio, luminosity and UV slopes, which have been reported in the literature.
BAL quasars, on average, have redder UV continua, higher Eddington ratios and luminosities than non-BAL
quasars. LoBAL quasars have more severe and slightly broader \ion{C}{4} absorption than HiBAL quasars.
Outflow velocities and strength are found to be moderately dependent on Eddington ratio and weakly
dependent on luminosity and UV slope. All of these results are consistent with previous works.

We find moderate and significant dependencies of outflow velocities and strength on $\beta_{\rm NIR}$,
in the sense that outflows strengthen with increasing hot dust emission. It is consistent with
the correlation of the blueshift of \ion{C}{4} BELs with $\beta_{\rm NIR}$ in non-BAL quasars (see paper I).
The large scatter in the relationship may be partly due to the fact that the outflow velocities and strength
measured from absorptions are sensitive to the LOS structure. Our statistical analysis and composite spectra
study both reveal that outflow strength is more strongly correlated with $\beta_{\rm NIR}$ than Eddington ratio,
luminosity and UV slope. In particular, the composites reveal that the entire \ion{C}{4} absorption profile
shifts blueward and broadens as $\beta_{\rm NIR}$ increases, while Eddington ratio and UV slope only affect the high and
low velocity part of outflows, respectively.

We discuss several potential processes that are possibly responsible for the observed correlations of outflow properties on hot
dust emission. In a dust-free outflow and hydrostatic and optically thick torus scenario, we attempt to use inclination
effect and other observed relationships associated with outflows to interpret the correlations. However, our analysis does
not favor this scenario. We thus suggest that dust is intrinsic to outflows and may have a nontrivial contribution to the outflow acceleration.

\acknowledgments
We thank the anonymous referee for a helpful report that significantly improves this paper.
This work was supported by Chinese Natural Science Foundation (NSFC-11033007, 1203021), the ``973" project (2013CB834905), NCET-11-0879 and Chinese Polar Environment Comprehensive Investigation \& Assessment Programmes (CHINARE-2014-02-03).

Funding for the SDSS and SDSS-II has been provided by the Alfred P. Sloan Foundation, the Participating Institutions, the National Science Foundation, the U.S. Department of Energy, the National Aeronautics and Space Administration, the Japanese Monbukagakusho, the Max Planck Society, and the Higher Education Funding Council for England. The SDSS Web Site is http://www.sdss.org/.

The SDSS is managed by the Astrophysical Research Consortium for the Participating Institutions. The Participating Institutions are the American Museum of Natural History, Astrophysical Institute Potsdam, University of Basel, University of Cambridge, Case Western Reserve University, University of Chicago, Drexel University, Fermilab, the Institute for Advanced Study, the Japan Participation Group, Johns Hopkins University, the Joint Institute for Nuclear Astrophysics, the Kavli Institute for Particle Astrophysics and Cosmology, the Korean Scientist Group, the Chinese Academy of Sciences ( LAMOST ), Los Alamos National Laboratory, the Max-Planck-Institute for Astronomy ( MPIA), the Max-Planck-Institute for Astrophysics (MPA), New Mexico State University, Ohio State University, University of Pittsburgh, University of Portsmouth, Princeton University, the United States Naval Observatory, and the University of Washington.

\appendix

\section{ A: Identification of absorption troughs}

In this section, we introduce our automatic algorithm to identify BALs.
In the Milky Way interstellar medium (ISM) studies, dust extinction curves of diffuse clouds were extracted by comparing
a pair of stellar spectra of the same spectral type, one of which is reddened and the other unreddened
(Fitzpatrick \& Massa 2007 and references therein). This method is called the "pair-method".
Here a similar method, the "quasar spectrum pair method", is adopted to
compare the spectra of BAL quasars with other unabsorbed to identify potential absorption troughs.
Lots of works suggested that BAL quasars have the similar continua and emission line profiles with non-BAL quasars
(e.g., Weymann et al. 1991, Korista et al. 1993).  Theoretically, we can always find a unabsorbed quasar which spectrum
adequately resembles that of a given BAL quasar. The non-BAL quasar spectrum can be used as an unabsorbed model spectrum
(hereafter model spectrum) to characterize the intrinsic continuum and BELs of the BAL quasar. The absorption troughs, if present,
are then shown in the spectrum of the BAL quasar normalized by the model spectrum.

For a BAL candidate in consideration, we employ our new automatic algorithm to seek the model spectrum through the following three steps.
\begin{description}
\item[1. Construct a quasar template spectrum library ]
Four hundred quasar spectra in the redshift range of $0<z-z_0<0.05$ are randomly selected from the non-BAL quasar catalog (see Section 2) to build up the quasar template spectrum library for the given BAL quasar candidate. Here $z$ and $z_0$ are the redshifts of the non-BAL quasar and the given candidate.
Then these template spectra are smoothed by twice iteration B-spline analysis\footnote{http://spectro.princeton.edu/idlutils\_doc.html}.
Pixels with flux intensities smaller than $85\%$ of the smoothed spectra may have narrow absorption lines, and therefore are masked out from the following fit.
\item[2. Model the given BAL candidate spectrum ]
The template spectra are scaled and multiplied by the SMC extinction law with a free parameter E(B-V)
to fit the observed spectrum of the given BAL candidate in two-step fitting loops.
The fitting process is performed using the MPFIT package (Markwardt 2009).
The scale parameter and E(B-V) are determined by minimizing the $\chi^2$.
After the first loop, we mask out the potential broad absorption regions where the observed flux are lower than $85\%$ of the scaled and reddened template, and refit the spectrum. The overlap spectral regions between the candidate spectrum and the template spectra, except those masked, are used in the fitting process and the calculation of $\chi^2$.
\item[3. Choose the best unabsorbed model spectrum]
We rank the 400 model spectra in the increasing order of $\chi^2$ (calculated in the second loop of step 2), and take out the first 20 model spectra. We calculate the emission line chi-square, $\chi^2_{\rm EL}$, in the wavelength range of [1450, 1650]\AA, where the pixels are not masked. This parameter measures the similarity between the observation and the model in the unabsorbed \ion{C}{4} BEL region.
The best model spectrum is chosen by minimizing $\chi^2_{\rm EL}$ among the twenty model spectra. Therefore our best model spectrum has the minimum difference in the \ion{C}{4} BEL regions from the candidate's spectrum, and is also in good agreement with the observed broad band continuum.
\end{description}

\figurenum{A1}
\begin{figure}[tbp]
\epsscale{0.5} \plotone{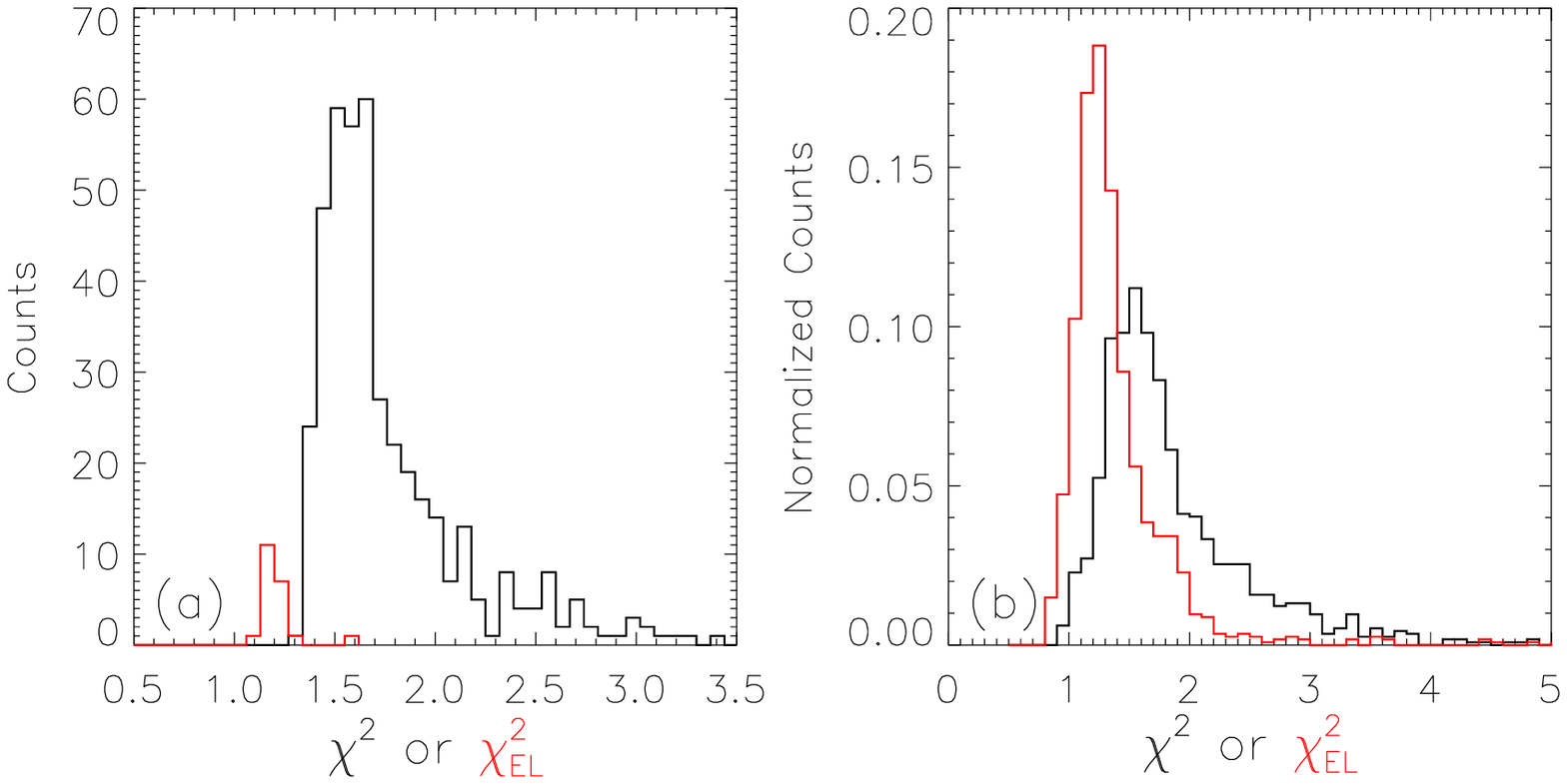}
\caption{Left panel: the number distributions of the reduced $\chi^2$ of the 400 model spectra for the BAL quasar candidate SDSS J010727.38-105743.1. The red histogram represents the distribution of the reduced $\chi^2$ and $\chi_{\rm EL}^2$ for the 20 smallest-$\chi^2$ model spectra. Right panel: the probability distributions of the reduced $\chi^2$ and $\chi_{\rm EL}^2$ for the chosen best model spectra of all candidate BAL quasars.
}\label{figa1}
\end{figure}

For illustration, we show, in Figure \ref{fig1}, the masked regions in gray for several candidate quasars. Almost all of the significant narrow absorption lines and the potential broad absorption lines are well identified. The quasar in the left bottom panel has low signal to noise ratio and many individual absorptions, so a considerable large part of the spectrum is masked.
In order to demonstrate the reliability of our algorithm, we show the probability distributions of the reduced $\chi^2$ and $\chi_{\rm EL}^2$ for the chosen best model spectra of all candidate BAL quasars in the right panel of Figure \ref{figa1}. The typical reduced $\chi^2$ is 1.5. Considering that we model the candidate spectra in a very large spectral region, our fitting results are quite good. We then show, in the left panel of Figure \ref{figa1}, the number count distribution of the reduced $\chi^2$ of 400 template spectra
which are used to model a typical candidate quasar SDSS J010727.38-105743.1. One can see that about half of template spectra are able to properly model the given spectra. And the distribution of the reduced $\chi_{\rm EL}^2$ for the 20 smallest-$\chi^2$ model spectra are shown in red. Obviously, the differences between the 20 model spectra and candidate spectrum in \civ\ BELs are also very small.

\section{B: Reliability of BAL identification}

To examine the reliability of our BAL quasar identification, we cross-match our working BAL quasar sample with the BAL quasar catalogs constructed by SC09, A11 and SH11. Note that the comparisons with SC09 and A11 are restricted to SDSS DR5 and DR6, respectively. Table A1 lists the numbers of BAL quasars identified or rejected by our work and the other three samples.
There are 1606 learning vector quantization (LVQ) BAL quasars, 1769 hybird-LVQ BAL quasars, 1315 A11's BAL quasars and 2395 SH11's BAL quasars which also belong to our $z\sim2$ DR7-WISE quasar sample (marked with ``T").
Most of them are classified as BAL quasars by this work (marked with ``B", see Table A1). Moreover we reject 284 LVQ BAL quasars, 134 hybird-LVQ BAL quasars, 65 A11's BAL quasars and 406 SH11's BAL quasars (marked with ``L"). In our working sample, there are 315, 2, 621 and 119 objects which are not included in LVQ, hybrid-LVQ, A11's and SH11's BAL quasar samples (marked with ``A"). The correctness and completeness of our working sample are very similar to the hybrid-LVQ BAL quasar sample identified by SC09.

\begin{figure}[tbp]
\epsscale{0.5} \plotone{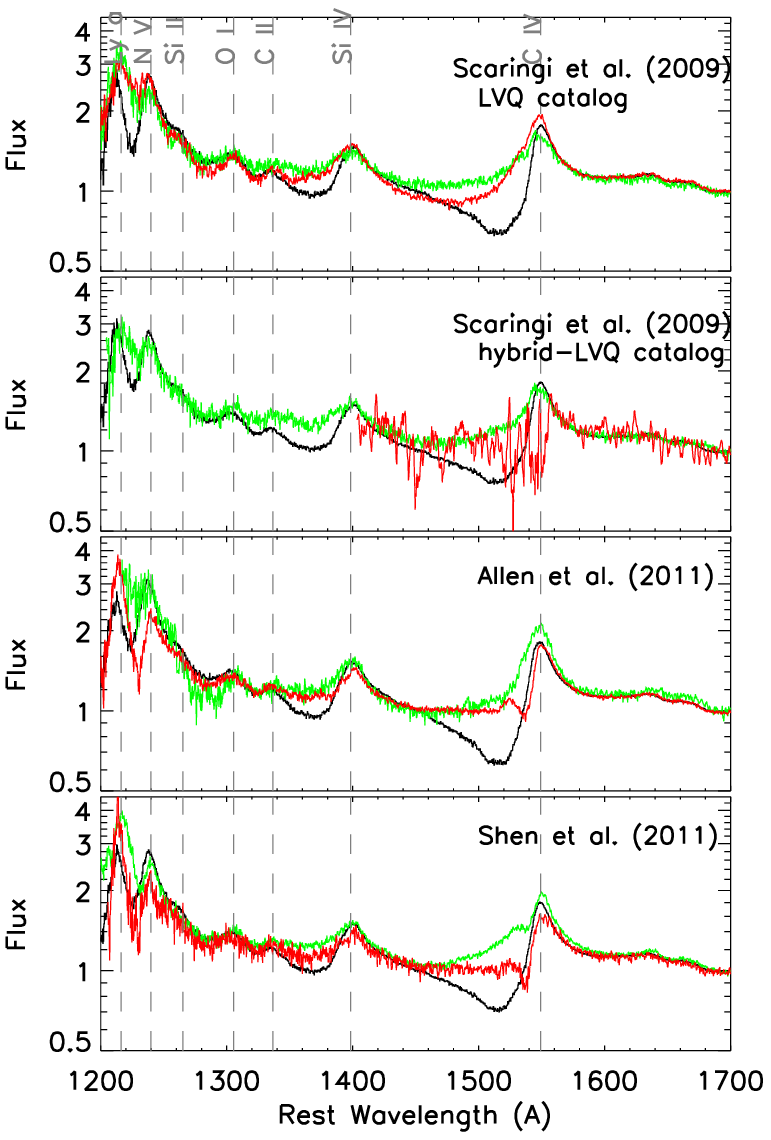}
\caption{Comparison of composite spectra of different BAL quasar subsamples. The black lines show the composite spectra
of BAL quasars which belong to our final working BAL quasar catalog and the BAL quasar catalog as indicated in the corresponding panel.
The green lines show the quasars which we rejected but are accepted by other works as indicated in the panel. And the red lines represent that we accepted  but are rejected by other works as indicated in the panel.
}\label{figa2}
\end{figure}

In Figure \ref{figa2}, we compare the composite spectra of the BAL quasar subsamples listed in Table A1.
The black, green and red lines represent the BAL quasar subsamples marked with ``B", ``L" and ``A", respectively.
One can see that there are almost no significant BALs in the composite spectra of the BAL candidates that we rejected but accepted by other works (green lines). On the contrary, the BALs are remarkable in the composites of BAL quasars we accepted but rejected by other works (red lines). A11 and SH11 tends to miss BAL quasars with low-velocity absorption.

\setcounter{table}{0}
\renewcommand\thetable{\Alph{table}\arabic{table}}
\begin{deluxetable*}{l cccc}
\tabletypesize{\scriptsize}
\tablecaption{The number of BAL quasars identified or rejected by various works.
\label{taba1} }
\tablewidth{0pt}
\startdata
\hline
 & DR5  & DR5   &   DR6 &   DR7\\
 \hline
  & Scaringi et al. (2009)  & Scaringi et al. (2009) & Allen et al. (2011)   &   Shen et al. (2011)\\
   & LVQ catalog  & hybrid-LVQ catalog &  BAL quasar catalog &BAL quasar catalog\\
   \hline
   T & 1606    &   1769&   1315    &   2395\\
   B & 1322    &   1635&   1250    &   1982\\
   L & 284     &   134 &   65      &   413 \\
   A & 315     &   2   &   621     &   117
   \enddata
   \tablenotetext{Note.} {T: BAL quasars in DR7-WISE quasar catalog which are identified by SC09, A11 and SH11.
   B: BAL quasars identified by this work and the other work listed in the second row.
   L: BAL quasars rejected by this work but accepted by the other work.
   A: BAL quasars accepted by this work but rejected by the other work.}
\end{deluxetable*}

\figurenum{A2}

\figurenum{A3}
\begin{figure}[tbp]
\epsscale{0.8} \plotone{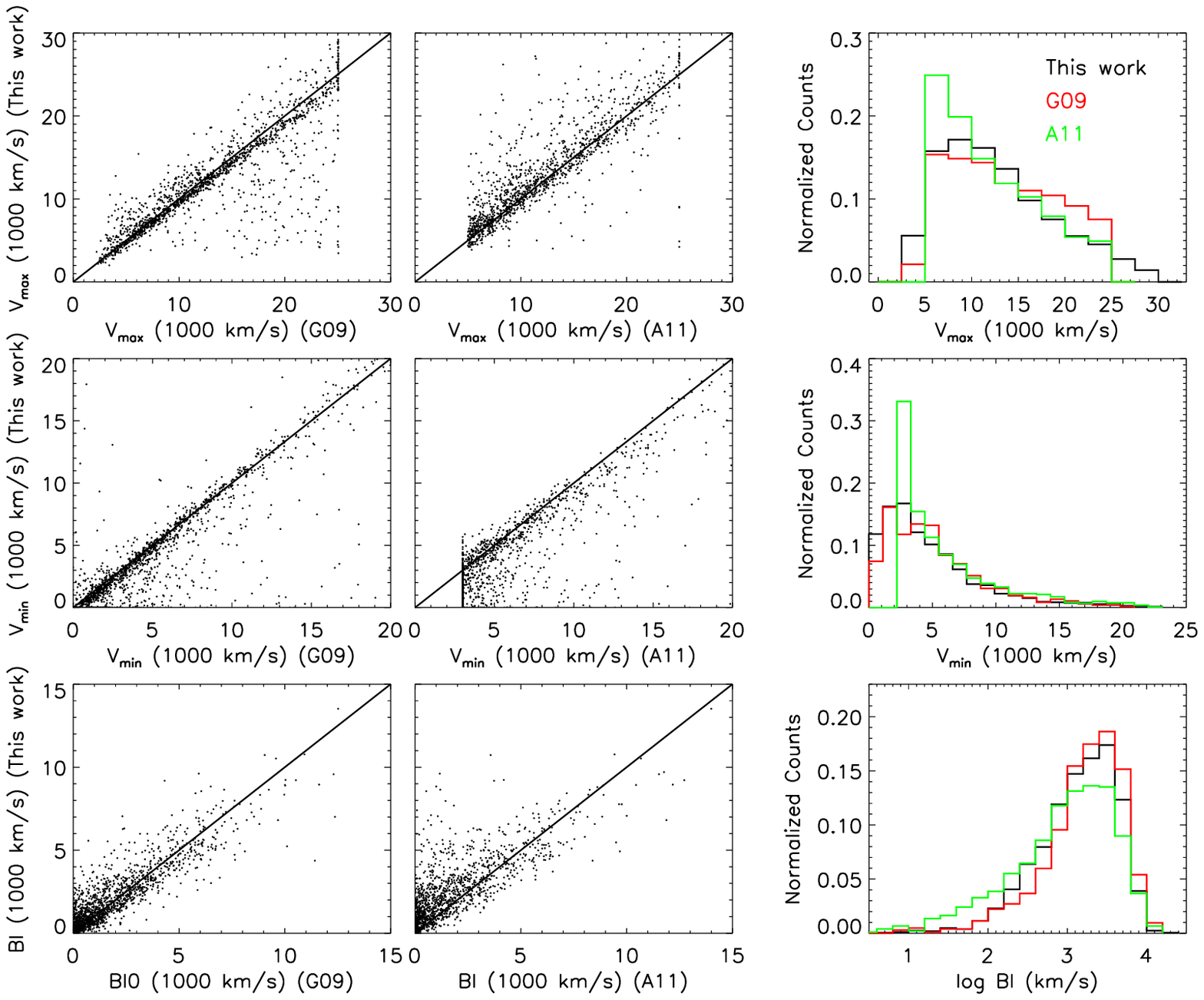}
\caption{Comparison of \ion{C}{4} BAL parameters presented in this paper, G09 and A11.
}\label{figa3}
\end{figure}

As shown in Table A1, there are 1250 objects which are included in both our and A11's samples. The BAL parameters are available for these
A11 objects. Moreover, among the 1982 BAL quasars included in both our and SH11's samples, 1517 objects have available BAL measurements from G09. Before presenting a comparison with these two works, we have to emphasize the differences in BAL identification among the three works, which may also have an impact on the BAL parameter measurements as shown below. As discussed in Section 2.1, $BI$ or its modifications are the most widely used metric to identify BALs, and all of the modifications follow the formula presented in Weymann et al. (1991). The only difference is in integration range. A11 adopted the traditional integration range from 3000 km/s to 25000 km/s. G09 extend to low-velocity end to zero in the calculation of $BI_0$. We adopt a even larger integration range from 0 km/s to 30000 km/s.

The direct comparison among these three works is shown in the left and middle panels of Figure \ref{figa3}. Our parameter measurements are well consistent with those of A11 and G09, although there is apparent discrepancy and scatter. Part of the discrepancies are due to the difference in BI definition as discussed above. For example, the high and low-velocity cutoffs in the $V_{\rm max}$ distribution of A11 sample are apparently ascribed to the chosen integration range. In addition, our $BI$ parameter tends to be greater than A11's at low-$BI$ end. It is also partly due to their smaller integration range than ours. The considerable scatter is also expected, since we use very different method to construct the model spectra.

In right panels, we show the parameter distributions of 1036 quasars, which are identified as BAL quasars by all the three works.
As discussed in Section 3.1, the $V_{max}$ distributions are different between ours and G09. Compared to our distribution, theirs is much flattened. However, our $V_{max}$ distribution well matches A11's, except at the high and low-velocity ends. The discrepancies at two ends are again caused by the different integration ranges. The three $V_{min}$ distributions are quite similar at $V_{min}>$ 4000 km/s. Since the starting velocity of BALs is set to 3000 km/s, A11's distribution has a significant high peak at $V_{\rm min}=3000$ km/s. The bottom
right panel shows that our $BI$ distribution is similar to G09's and A11' balnicity trends to smaller than ours because of the integration range.

Overall our comprehensive comparison suggests that our BAL quasar sample is at least as good as, or even better than, those in the literature.

\section{ C: Uncertainties of BAL parameters}

\figurenum{A4}
\begin{figure}[tbp]
\epsscale{0.45} \plotone{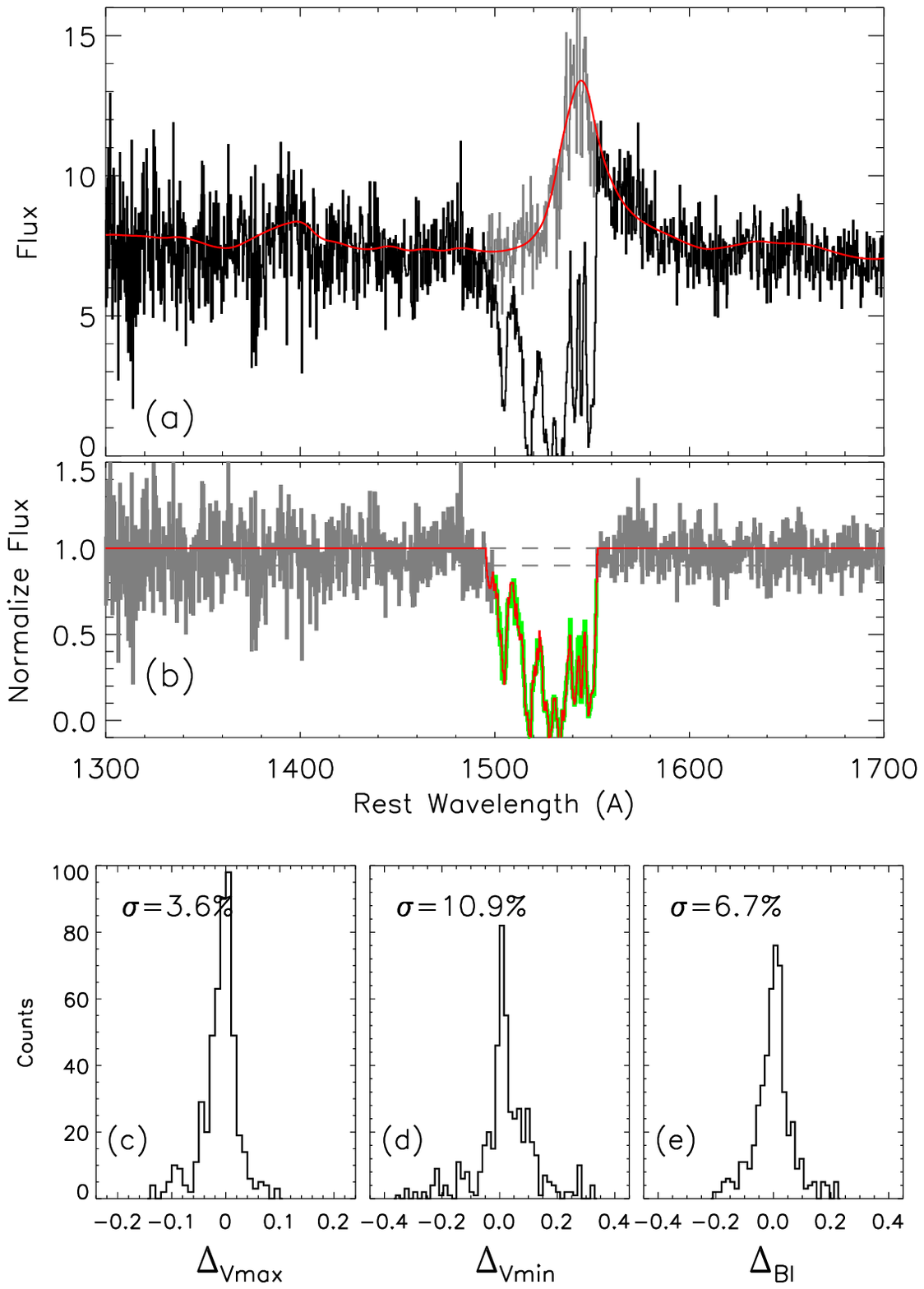}
\caption{Panel (a): Black line: simulated spectrum which is constructed from the observed spectrum (gray)
of SDSS J122626.22+270437.0 multiplied by the \ion{C}{4} absorption profile obtained from SDSS J035310.23+094418.7.
Red line: the best model spectrum for this simulated spectrum.
Panel (b): Red line: the input \ion{C}{4} absorption profile of SDSS J035310.23+094418.7. Green line: the absorption profile identified from the simulated spectrum. Gray line: the normalized flux of the simulated spectrum.
Panel (c-e): the distributions of the relative difference $\Delta_x$  between the input parameters and the parameters measured from the simulated spectra. Here $\Delta_x = (x_{\rm input}-x_{\rm sim})/x_{\rm input}$, $x$ in the three panels denotes $V_{\rm max}$, $V_{\rm min}$ and $BI$ respectively.}\label{figa4}
\end{figure}

The measurement uncertainties of BAL parameters arise from the statistical noise and the uncertainties in model spectrum.
The measurement uncertainty due to the statistical noise is usually easily to handle. However, the model uncertainties is hard estimated.
We therefore construct 500 simulated BAL quasar spectra to quantify the uncertainties. The simulated spectra are
constructed via multiplying the randomly selected non-BAL quasar spectra by absorption profiles. To mimic BALs in a realistic way, the absorption profiles are also extracted from randomly selected BAL quasars. Then we fit the simulated spectra and measure the BAL parameters following our automatic algorithm. We emphasize that the non-BAL spectra selected to create simulated spectra are rejected during the fitting process.

As a simple demonstration, we show, in Figure \ref{figa4}, the non-BAL quasar spectrum of SDSS J122626.22+270437.0 (gray curve in panel (a)), the observed \ion{C}{4} absorption profile of SDSS J035310.23+094418.7 (red line in panel (b)), and the simulated BAL quasar spectrum (black curve in panel (a)). The best model spectrum for the simulated spectrum is shown in red line in panel (a) and the normalized flux in gray line in panel (b). The identified absorption profile (green line in panel (b)) is very similar to the input profile, despite of tiny difference. One can see that the maximum velocity is reduced because of the flux noises. And the absorption close to 1549\AA~ is slightly shallower than the input one, which is likely caused by the BEL profile mismatch.

We have original $V_{\rm max}$, $V_{\rm min}$ and $BI$ for the 500 simulated spectra. These original quantities, $X_{\rm input}$, are directly measured from the input profiles. We also get the corresponding parameters from the simulated spectra, $X_{\rm sim}$, using our automatic algorithm. So we can derive the uncertainties in the three parameters using the formula $\Delta_x = (x_{\rm input}-x_{\rm sim})/x_{\rm input}$. The distributions of $\Delta_x$ are shown in panel (c-e) of Figure \ref{figa4}. The standard deviations of the $\Delta_x$ distributions can be used to represent the typical $1\sigma$ errors. The $1\sigma$ errors for $V_{\rm max}$, $V_{\rm min}$ and $BI$ are 3.6\%, 10.9\% and 6.7\%, respectively. We note that the resultant errors contain both the uncertainties due to statistical noise and the model uncertainty.

\end{document}